\documentclass[12pt]{elsart} %

\usepackage{graphicx}
\usepackage{amsmath}
\usepackage{epsfig}
\newcommand{\ksbarz}{\ensuremath{\overline{K}{}^\ast(892)^0}}
\newcommand{\mev}{\ensuremath{\mathrm{MeV}}}
\newcommand{\gev}{\ensuremath{\mathrm{GeV}}}
\newcommand{\like}{\ensuremath{\mathcal{L}}}
\newcommand{\lrat}{\ensuremath{\mathcal{P}}}

\begin{document}

\begin{frontmatter}
\epsfysize3cm

\hspace{-9.5cm}
\epsfbox{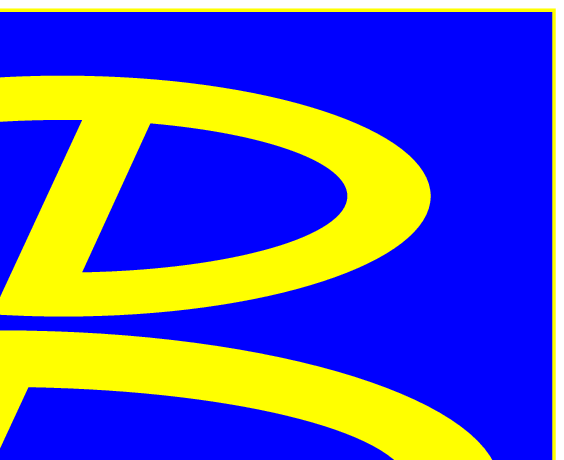}    % BELLE-logo
\begin{flushright}
\vskip -3cm
\noindent
\hspace*{9.0cm}BELLE Preprint 2004-27\\
\hspace*{9.0cm}KEK Preprint 2004-54\\
\hspace*{9.0cm}\today\\
%\hspace*{9.0cm}version 15\\
\end{flushright}                   
\normalfont

\vspace*{2.5cm}

\title{\Large
Measurement of masses and branching ratios of~$\Xi_c^+$ 
and~$\Xi_c^0$ baryons} 
\baselineskip=.3cm

%%%%%%%%%%%%%%%%%%%%%%%%%%%%%%%%%%%%%%%%%%%%%%%%%%%%%%%%%%%%%%%%%%%%%
%\input{author}
%%% Journal:  Physics Letters B
%%% Contacts: T. Lesiak (tadeusz.lesiak@ifj.edu.pl)
%%%           B. Yabsley (yabsley@bmail.kek.jp)
%%% Paper:   xi_c baryons
%%% Non-responding authors or those who said NO are commented out.
%%% ====================================================================
%%% Click the RELOAD button on your web browser to see the updated file.
%%% ====================================================================
%%% Use \input{author} to insert this material into your latex file.
\collab{Belle Collaboration}
  \author[Krakow]{T.~Lesiak}, % Krakow
  \author[KEK]{K.~Abe}, % KEK
  \author[TohokuGakuin]{K.~Abe}, % TohokuGakuin
% \author[TIT]{N.~Abe}, % TIT
  \author[KEK]{I.~Adachi}, % KEK
  \author[Tokyo]{H.~Aihara}, % Tokyo
% \author[Nagoya]{M.~Akatsu}, % Nagoya
  \author[Tsukuba]{Y.~Asano}, % Tsukuba
% \author[Toyama]{T.~Aso}, % Toyama
% \author[BINP]{V.~Aulchenko}, % BINP
% \author[ITEP]{T.~Aushev}, % ITEP
% \author[Tata]{T.~Aziz}, % Tata
  \author[Cincinnati]{S.~Bahinipati}, % Cincinnati
  \author[Sydney]{A.~M.~Bakich}, % Sydney
% \author[Peking]{Y.~Ban}, % Peking
% \author[Hawaii]{M.~Barbero}, % Hawaii
  \author[Lausanne]{A.~Bay}, % Lausanne
% \author[BINP]{I.~Bedny}, % BINP
  \author[JSI]{U.~Bitenc}, % Ljubljana
% ...
$^{,\dag}$, % Tsukuba
  \author[JSI]{M.~Stari\v c}, % Ljubljana
% \author[Nagoya]{A.~Sugi}, % Nagoya
% \author[Saga]{A.~Sugiyama}, % Saga
% \author[Osaka]{K.~Sumisawa}, % Osaka
  \author[TMU]{T.~Sumiyoshi}, % TMU
% \author[Saga]{S.~Suzuki}, % Saga
% \author[KEK]{S.~Y.~Suzuki}, % KEK
% \author[Hawaii]{S.~K.~Swain}, % Hawaii
% \author[KEK]{O.~Tajima}, % KEK
% \author[KEK]{F.~Takasaki}, % KEK
  \author[KEK]{K.~Tamai}, % KEK
  \author[Niigata]{N.~Tamura}, % Niigata
% \author[Tokyo]{K.~Tanabe}, % Tokyo
  \author[KEK]{M.~Tanaka}, % KEK
% \author[Melbourne]{G.~N.~Taylor}, % Melbourne
  \author[OsakaCity]{Y.~Teramoto}, % OsakaCity
% \author[Peking]{X.~C.~Tian}, % Peking
% \author[Nagoya]{S.~Tokuda}, % Nagoya
% \author[Melbourne]{S.~N.~Tovey}, % Melbourne
% \author[Hawaii]{K.~Trabelsi}, % Hawaii
  \author[KEK]{T.~Tsuboyama}, % KEK
  \author[KEK]{T.~Tsukamoto}, % KEK
% \author[Hawaii]{K.~Uchida}, % Hawaii
  \author[KEK]{S.~Uehara}, % KEK
  \author[Taiwan]{K.~Ueno}, % Taiwan
% \author[ITEP]{T.~Uglov}, % ITEP
% \author[Chiba]{Y.~Unno}, % Chiba
  \author[KEK]{S.~Uno}, % KEK
  \author[KEK]{Y.~Ushiroda}, % KEK
  \author[Hawaii]{G.~Varner}, % Hawaii
  \author[Sydney]{K.~E.~Varvell}, % Sydney
  \author[Lausanne]{S.~Villa}, % Lausanne
  \author[Taiwan]{C.~C.~Wang}, % Taiwan
  \author[Lien-Ho]{C.~H.~Wang}, % Lien-Ho
% \author[VPI]{J.~G.~Wang}, % VPI
% \author[Taiwan]{M.-Z.~Wang}, % Taiwan
% \author[Niigata]{M.~Watanabe}, % Niigata
% \author[TIT]{Y.~Watanabe}, % TIT
% \author[Vienna]{L.~Widhalm}, % Vienna
  \author[IHEP]{Q.~L.~Xie}, % IHEP
  \author[VPI]{B.~D.~Yabsley}, % VPI
  \author[Tohoku]{A.~Yamaguchi}, % Tohoku
% \author[Tohoku]{H.~Yamamoto}, % Tohoku
% \author[TMU]{S.~Yamamoto}, % TMU
% \author[Osaka]{T.~Yamanaka}, % Osaka
  \author[NihonDental]{Y.~Yamashita}, % NihonDental
% \author[KEK]{M.~Yamauchi}, % KEK
% \author[Seoul]{Heyoung~Yang}, % Seoul
% \author[Taiwan]{P.~Yeh}, % Taiwan
  \author[Peking]{J.~Ying}, % Peking
% \author[Nagoya]{K.~Yoshida}, % Nagoya
% \author[IHEP]{Y.~Yuan}, % IHEP
  \author[Tohoku]{Y.~Yusa}, % Tohoku
% \author[Aomori]{H.~Yuta}, % Aomori
% \author[IHEP]{S.~L.~Zang}, % IHEP
  \author[IHEP]{C.~C.~Zhang}, % IHEP
  \author[KEK]{J.~Zhang}, % KEK
  \author[USTC]{L.~M.~Zhang}, % USTC
  \author[USTC]{Z.~P.~Zhang}, % USTC
% \author[Hawaii]{Y.~Zheng}, % Hawaii
  \author[BINP]{V.~Zhilich}, % BINP
% \author[Princeton]{T.~Ziegler}, % Princeton
and
  \author[Ljubljana,JSI]{D.~\v Zontar} % Ljubljana
% \author[Lausanne]{D.~Z\"urcher}, % Lausanne

\address[Aomori]{Aomori University, Aomori, Japan}
\address[BINP]{Budker Institute of Nuclear Physics, Novosibirsk, Russia}
\address[Chiba]{Chiba University, Chiba, Japan}
\address[Chonnam]{Chonnam National University, Kwangju, South Korea}
\address[Chuo]{Chuo University, Tokyo, Japan}
\address[Cincinnati]{University of Cincinnati, Cincinnati, OH, USA}
\address[Frankfurt]{University of Frankfurt, Frankfurt, Germany}
\address[Gyeongsang]{Gyeongsang National University, Chinju, South Korea}
\address[Hawaii]{University of Hawaii, Honolulu, HI, USA}
\address[KEK]{High Energy Accelerator Research Organization (KEK), Tsukuba, Japan}
\address[Hiroshima]{Hiroshima Institute of Technology, Hiroshima, Japan}
\address[IHEP]{Institute of High Energy Physics, Chinese Academy of Sciences, Beijing, PR China}
\address[Vienna]{Institute of High Energy Physics, Vienna, Austria}
\address[ITEP]{Institute for Theoretical and Experimental Physics, Moscow, Russia}
\address[JSI]{J. Stefan Institute, Ljubljana, Slovenia}
\address[Kanagawa]{Kanagawa University, Yokohama, Japan}
\address[Korea]{Korea University, Seoul, South Korea}
\address[Kyoto]{Kyoto University, Kyoto, Japan}
\address[Kyungpook]{Kyungpook National University, Taegu, South Korea}
\address[Lausanne]{Swiss Federal Institute of Technology of Lausanne, EPFL, Lausanne}
\address[Ljubljana]{University of Ljubljana, Ljubljana, Slovenia}
\address[Maribor]{University of Maribor, Maribor, Slovenia}
\address[Melbourne]{University of Melbourne, Victoria, Australia}
\address[Nagoya]{Nagoya University, Nagoya, Japan}
\address[Nara]{Nara Women's University, Nara, Japan}
\address[NCU]{National Central University, Chung-li, Taiwan}
\address[Kaohsiung]{National Kaohsiung Normal University, Kaohsiung, Taiwan}
\address[Lien-Ho]{National United University, Miao Li, Taiwan}
\address[Taiwan]{Department of Physics, National Taiwan University, Taipei, Taiwan}
\address[Krakow]{H. Niewodniczanski Institute of Nuclear Physics, Krakow, Poland}
\address[NihonDental]{Nihon Dental College, Niigata, Japan}
\address[Niigata]{Niigata University, Niigata, Japan}
\address[OsakaCity]{Osaka City University, Osaka, Japan}
\address[Osaka]{Osaka University, Osaka, Japan}
\address[Panjab]{Panjab University, Chandigarh, India}
\address[Peking]{Peking University, Beijing, PR China}
\address[Princeton]{Princeton University, Princeton, NJ, USA}
\address[RIKEN]{RIKEN BNL Research Center, Brookhaven, NY, USA}
\address[Saga]{Saga University, Saga, Japan}
\address[USTC]{University of Science and Technology of China, Hefei, PR China}
\address[Seoul]{Seoul National University, Seoul, South Korea}
\address[Sungkyunkwan]{Sungkyunkwan University, Suwon, South Korea}
\address[Sydney]{University of Sydney, Sydney, NSW, Australia}
\address[Tata]{Tata Institute of Fundamental Research, Bombay, India}
\address[Toho]{Toho University, Funabashi, Japan}
\address[TohokuGakuin]{Tohoku Gakuin University, Tagajo, Japan}
\address[Tohoku]{Tohoku University, Sendai, Japan}
\address[Tokyo]{Department of Physics, University of Tokyo, Tokyo, Japan}
\address[TIT]{Tokyo Institute of Technology, Tokyo, Japan}
\address[TMU]{Tokyo Metropolitan University, Tokyo, Japan}
\address[TUAT]{Tokyo University of Agriculture and Technology, Tokyo, Japan}
\address[Toyama]{Toyama National College of Maritime Technology, Toyama, Japan}
\address[Tsukuba]{University of Tsukuba, Tsukuba, Japan}
\address[Utkal]{Utkal University, Bhubaneswer, India}
\address[VPI]{Virginia Polytechnic Institute and State University, Blacksburg, VA, USA}
\address[Yonsei]{Yonsei University, Seoul, South Korea}
%%%%%%%%%%%%%%%%%%%%%%%%%%%%%%%%%%%%%%%%%%%%%%%%%%%%%%%%%%%%%%%%%%%%%%%%%%%%%%%%%%
\vfill

%\thanks[NovaGorica]{on leave from Nova Gorica Polytechnic, Nova Gorica, Slovenia}
{\raggedleft $^\dag$ {\small on leave from Nova Gorica Polytechnic, Nova Gorica, Slovenia}}
\clearpage

%\vfill

\baselineskip=4truept
\overfullrule=0truept
\leftskip=0.5truein
\rightskip=0.5truein  
\small

\newpage

\begin{center}
\begin{abstract}
We report a measurement of the $\Xi_c^+$ and $\Xi_c^0$ baryon masses,
and the branching ratios for various $\Xi_c$ decays, 
using $140\,\mathrm{fb}^{-1}$ of data collected by the Belle experiment
at the KEKB $e^+ e^-$ collider.
The mass splitting $m_{\Xi_c^0} - m_{\Xi_c^+}$ is found to be
$2.9\pm 0.5\,\mathrm{MeV}/c^2$; this measurement is three times as precise
as the current world average.
We measure the branching ratios
$\Gamma (\Xi_c^+\to \Lambda K\pi\pi)/\Gamma (\Xi_c^+\to\Xi\pi\pi) = 0.32\pm 0.03 \pm 0.02$ and  
$\Gamma (\Xi_c^0\to  p KK\pi)/\Gamma (\Xi_c^0\to\Xi\pi)         = 0.33\pm 0.03 \pm 0.03$, 
with improved precision, 
and measure 
$\Gamma (\Xi_c^+\to p  K^0_S K^0_S)/\Gamma (\Xi_c^+\to\Xi\pi\pi)     = 0.087\pm 0.016 \pm 0.014$,  
$\Gamma (\Xi_c^0\to \Lambda K\pi)/\Gamma (\Xi_c^0\to\Xi\pi)        = 1.07\pm 0.12 \pm 0.07$ and  
$\Gamma (\Xi_c^0\to \Lambda K^0_S)/\Gamma (\Xi_c^0\to\Xi\pi)       = 0.21\pm 0.02 \pm 0.02$  
for the first time. 
In $\Xi_c^0$ decays to the $p K^- K^- \pi^+$ final state, 
we find evidence for the process $\Xi_c^0 \to p K^- \ksbarz$
and measure the fraction of decays via this process to be $0.51\pm 0.03 \pm 0.01$.
\par
\vskip 3mm
{\bf Author Keywords} Charmed baryon; W-exchange
\vskip 3mm
{\bf PACS classification codes} 13.30Eg; 14.20Lq
\end{abstract}
\end{center}

\end{frontmatter}

%\leftskip=0truept
%\rightskip=0truept
\normalsize
\vspace{0.13in}

%%%%%%%%%%%%%%%%%%%%%%%%%%%%%%%%%%%%%%%%%%%%%%%%%%%%%%%%%%%%%%%%%%%%%%%%%%%%%%

\section{Introduction}

%%%%%%%%%%%%%%%%%%%%%%%%%%%%%%%%%%%%%%%%%%%%%%%%%%%%%%%%%%%%%%%%%%%%%%%%%%%%%%

Despite significant progress in experimental studies of charmed 
baryons, the properties of the $\Xi_c$ baryons are still poorly known. 
The current world average masses are
$(2466.3\pm 1.4)\,\mev/c^2$ for the~$\Xi_c^+$ and
$(2471.8\pm 1.4)\,\mev/c^2$ for the~$\Xi_c^0$,
and the precision on the mass splitting is comparable,
$\pm 1.8\,\mev/c^2$~\cite{PDG}.  
Among the exclusive decays reported so far,
only the observations by the CLEO~\cite{CLEO96a,CLEO96b,CLEO03} 
and FOCUS~\cite{FOCUS01,FOCUS03} collaborations
are based on data samples of more than 100 events. 
No absolute branching fractions have been measured, 
and branching ratios relative to the `reference modes'
$\Xi_c^+ \to \Xi^-\pi^+\pi^+$ and $\Xi_c^0 \to \Xi^-\pi^+$
have been determined with a typical precision of only 30\%.  

This Letter presents the results of a study of exclusive $\Xi_c$ decays 
in $e^+e^-$ continuum production,
with $\approx 3000$ observed events in the reference modes.
Branching ratios for $\Xi_c^+$ decays to the
$\Lambda K^- \pi^+\pi^+$ and $p K^0_S K^0_S$ final states,\footnote{
	Charge conjugate modes are included everywhere, 
	unless otherwise specified.}  
and for $\Xi^0_c$ decays to 
$\Lambda K^- \pi^+$, $\Lambda K^0_S$ and $p K^- K^- \pi^+$,
have been measured with a typical precision of $\approx 15\%$.  
The large reconstructed samples also allow precise measurements
of the $\Xi_c$ masses, and in particular the mass splitting
between the neutral and charged states.

\begin{figure}[hbt]\centering
\setlength{\unitlength}{1mm}
\begin{picture}(150,70)
\thicklines
%\put(0,0){\framebox(110,65){}}
%----------------------------------------
\put(10,60){\Large\bf a)}
\put( 0,46){\line(1,0){60}}
\put( 7,46){\vector(1,0){3}}
\put(46,46){\vector(1,0){3}}
\put( 3,48){\large\bf d}
\put(42,48){\large\bf u}
%----------------------------------------
\put( 0,30){\line(1,0){60}}
\put( 7,30){\vector(1,0){3}}
\put(46,30){\vector(1,0){3}}
\put( 3,32){\large\bf c}
\put(42,32){\large\bf s}
%----------------------------------------
\multiput(29,31)(0,4){4}{\line(0,1){2}}
%\put(29,30){\line(0,1){15}}
\put(22,35){\large\bf W}
%----------------------------------------
\put(-5,21){\large\bf $\Xi_c^0$}
%----------------------------------------
\put(40,19){\line(4,1){20}}
\put(40,19){\line(4,-1){20}}
\put(48,17){\vector(-3,1){2}}
\put(47,20.7){\vector(3,1){2}}
\put(42,23){\large\bf d}
\put(42,12){\large\bf $\overline{\mathrm d}$}
%----------------------------------------
\put( 0, 6){\line(1,0){60}}
\put( 7, 6){\vector(1,0){3}}
\put(46, 6){\vector(1,0){3}}
\put( 3, 8){\large\bf s}
\put(42, 8){\large\bf s}
%----------------------------------------
\put(62,32){\large\bf $\Lambda^0$}
\put(62, 8){\large\bf $\overline{K}{}^0$}
%----------------------------------------
%----------------------------------------
%----------------------------------------
\put(95,60){\Large\bf b)}
\put( 85,46){\line(1,0){60}}
\put( 92,46){\vector(1,0){3}}
\put(131,46){\vector(1,0){3}}
\put( 88,48){\large\bf d}
\put(127,48){\large\bf d}
%----------------------------------------
\put( 85,30){\line(1,0){60}}
\put( 92,30){\vector(1,0){3}}
\put(131,30){\vector(1,0){3}}
\put( 88,32){\large\bf c}
\put(127,32){\large\bf s}
%----------------------------------------
%\put(125,30){\line(0,-1){11}}
\multiput(125,30)(0,-4){3}{\line(0,-1){2}}
\put(117,22){\large\bf W}
%----------------------------------------
\put(80,21){\large\bf $\Xi_c^0$}
%----------------------------------------
\put(125,19){\line(4,1){20}}
\put(125,19){\line(4,-1){20}}
\put(133,17){\vector(-3,1){2}}
\put(132,20.7){\vector(3,1){2}}
\put(127,23){\large\bf u}
\put(127,12){\large\bf $\overline{\mathrm d}$}
%----------------------------------------
\put( 85, 6){\line(1,0){60}}
\put( 92, 6){\vector(1,0){3}}
\put(131, 6){\vector(1,0){3}}
\put( 88, 8){\large\bf s}
\put(127, 8){\large\bf s}
%----------------------------------------
\put(147,32){\large\bf $\Lambda^0$}
\put(147, 8){\large\bf $\overline{K}{}^0$}
%----------------------------------------
\end{picture}
\caption{Feynman diagrams for the process $\Xi_c^0\to \Lambda K^0_S$:
 ({\bf a}) W exchange and ({\bf b})
internal spectator.}
\label{DIAGL0K0}
\end{figure}
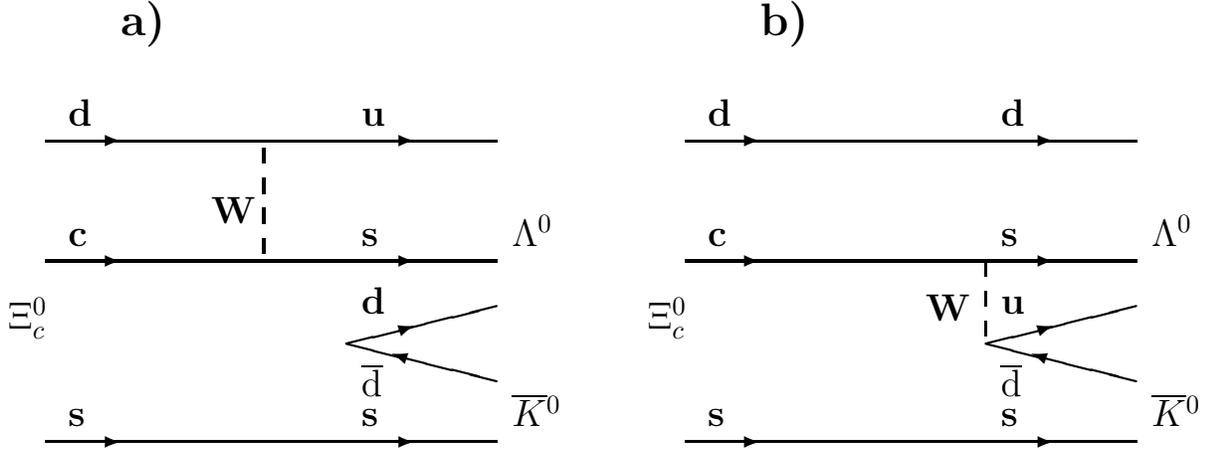

The decay $\Xi_c^0\to \Lambda K^0_S$ is of particular interest,
as it can occur only via
the poorly known $W$-boson-exchange process (Fig.~\ref{DIAGL0K0}(a)) or
the internal-spectator diagram (Fig.~\ref{DIAGL0K0}(b)),
in the absence of final state interactions.
Theoretical predictions for this mode are based, for example, on 
a symmetric-quark-model approach~\cite{SYMQUA1,SYMQUA2},
and span a range of branching fractions from 0.4\% to 0.7\%~\cite{ZENCZYK,SHARMA}; 
the fraction for the reference decay $\Xi_c\to \Xi^-\pi^+ $
is predicted to lie between 0.9\% and 2\%.

This paper is organized as follows. Section~\ref{SAMPLE} describes the detector
and data sample, and Section~\ref{RECON} describes the reconstruction
of~$\Xi_c$ baryons. The remaining sections present our determination of the
$\Xi_c$ masses      (Section~\ref{MASS}) and
branching fractions (Section~\ref{BRANCHING}).
Section~\ref{SUBSTRUCTURE} presents a study of the resonant substructure of
the $\Xi_c^0\to p K^- K^- \pi^+$ decay, improving on the precision of the
recent CLEO measurement~\cite{CLEO03}.

%%%%%%%%%%%%%%%%%%%%%%%%%%%%%%%%%%%%%%%%%%%%%%%%%%%%%%%%%%%%%%%%%%%
 
\section{Detector and data sample}
\label{SAMPLE}

The data used for this study were collected on the $\Upsilon(4S)$ resonance
using the Belle detector at the KEKB asymmetric $e^+e^-$ collider~\cite{KEKB}.
The integrated luminosity of the data sample is~140~fb$^{-1}$.

The Belle detector is a large-solid-angle magnetic spectrometer that consists of 
a~three-layer silicon vertex detector (SVD), a 50-layer central drift chamber (CDC),
an array of aerogel threshold \v{C}erenkov counters (ACC),
 a barrel-like arrangement of 
time-of-flight scintillation counters (TOF), and an electromagnetic calorimeter 
comprised of CsI(Tl) crystals (ECL) located inside a super-conducting solenoid coil 
that provides a 1.5~T magnetic field. An iron flux-return located outside of the  
coil is instrumented to detect $K^0_L$ mesons and to identify muons (KLM). 
A detailed  description of the Belle detector can be found elsewhere~\cite{BELLE}.

%%%%%%%%%%%%%%%%%%%%%%%%%%%%%%%%%%%%%%%%%%%%%%%%%%%%%%%%%%%%%%%%%%%%%%%%%%%%%%

\section{Reconstruction}
\label{RECON}

%%%%%%%%%%%%%%%%%%%%%%%%%%%%%%%%%%%%%%%%%%%%%%%%%%%%%%%%%%%%%%%%%%%%%%%%%%%%%%

Reconstruction of $\Xi_c$ decays for this analysis proceeds in three steps:
reconstruction of tracks and their identification as protons, kaons or pions;
combination of tracks to reconstruct $K^0_S$ mesons and $\Lambda$ and $\Xi^-$
hyperons;
and the selection of $\Xi_c$ candidates from combinations of tracks, $K^0_S$'s
and hyperons.
The method for each step is described in the following sections in turn.

%%%%%%%%%%%%%%%%%%%%%%%%%%%%%%%%%%%%%%%%%%%%%%%%%%%%%%%%%%%%%%%%%%%%%%%%%%%%%%

\subsection{Track reconstruction and identification}

Charged tracks are reconstructed from hits in the CDC using a Kalman filter~\cite{KALMAN},
and matched to hits in the SVD where present.
Quality criteria are then applied.
Excepting those tracks used to form $K^0_S$, $\Lambda$ and $\Xi^-$ candidates,
all tracks are required to have impact parameters relative to the
interaction point (IP) of less than 0.5~cm in the $r-\phi$ plane,
and 5~cm in the $z$ direction. (The $z$-axis is oriented opposite to the 
direction of the $e^+$ beam, along the symmetry axis of the detector.)
The transverse momentum of each track is required to exceed $0.1\,\gev/c$,
in order to reduce the low momentum combinatorial background.

Identification of tracks is based on information from
the CDC (energy loss $dE/dx$), TOF and ACC, combined to form likelihoods 
$\like(p)$, $\like(K)$ and $\like(\pi)$ for the proton, kaon and
pion hypotheses respectively.
These likelihoods are combined to form ratios
$\lrat(K/\pi) = \like(K) / (\like(K) + \like(\pi))$
and $\lrat(p/K) = \like(p) / (\like(p) + \like(K))$,
spanning the range from zero to one, which are then used to select track
samples.
Kaon candidates are required to satisfy $\lrat(K/\pi) > 0.9$ and
$\lrat(p/K) < 0.98$; the second criterion is to veto protons.
This selection has an efficiency of 80\% and a fake rate of 3.8\% ($\pi$ fakes $K$).
Protons are required to satisfy $\lrat(p/K) > 0.9$.
Pion candidates, except those coming from the decay of the 
$\Lambda$ hyperon, should satisfy both a proton and a kaon veto: 
$\lrat(p/K) < 0.98$ and $\lrat(K/\pi) < 0.98$. 

Electrons are identified using a similar likelihood ratio
$\lrat_e = \like_e / (\like_e + \like_{\text{non-}e})$,
based on a combination of $dE/dx$ measurements in the CDC, 
the response of the ACC, and $E/p$, where $p$ is the momentum of the track
and $E$ the energy of the associated cluster in the ECL. 
All tracks with $\lrat_e > 0.98$ are assumed to be electrons,
and removed from the proton, kaon and pion samples.

%%%%%%%%%%%%%%%%%%%%%%%%%%%%%%%%%%%%%%%%%%%%%%%%%%%%%%%%%%%%%%%%%%%%%%%%%%%%%%

\subsection{Reconstruction of $\Lambda$, $K^0_S$, and $\Xi^-$}

We reconstruct $\Lambda$ hyperons in the $\Lambda\to p\pi$ decay mode,
requiring the proton track to satisfy $\lrat(p/K) > 0.1$, and fitting the
$p$ and $\pi$ tracks to a common vertex.
The $\chi^2/n.d.f.$ of the vertex should not exceed 25, and the difference in the $z$-coordinate between the proton and pion at the 
vertex is required to be less than 2~cm. Due to the large $c\tau$ factor for $\Lambda$
hyperons (7.89~cm), we demand that the distance between the decay vertex and 
the IP in the $r-\phi$ plane be greater than 1~cm.
The invariant mass of the proton-pion pair is required to be within
$2.4\,\mev/c^2$ ($\approx 2.5$ standard deviations) of the nominal $\Lambda$ mass.

$K^0_S$ mesons are reconstructed using pairs of charged tracks that have an 
invariant mass  within $6\,\mev/c^2$ (2.5 standard deviations)
of the nominal $K^0_S$ mass, and a well
reconstructed vertex displaced from the IP by at least 5~mm.

We reconstruct $\Xi^-$ hyperons in the decay mode $\Xi^-\to \Lambda\pi^-$.
The $\Lambda$ and~$\pi$ candidates are fitted to a common vertex, 
whose $\chi^2/n.d.f.$ is required to be at most~25.
The distance between the $\Xi^-$ vertex position and interaction point in the 
$r-\phi$ plane should be at least 5~mm, and
less than the corresponding distance between the IP and the $\Lambda$ vertex.
The invariant mass of the $\Lambda \pi^-$ pair is required to be within
$7.5\,\mev/c^2$ of the nominal value ($\approx 2.5$ standard deviations).

%%%%%%%%%%%%%%%%%%%%%%%%%%%%%%%%%%%%%%%%%%%%%%%%%%%%%%%%%%%%%%%%%%%%%%%%%%%%%%

\subsection{Reconstruction of $\Xi_c^+$ and $\Xi_c^0$}

Charged hadrons, $K^0_S$ mesons and $\Lambda$ and $\Xi^-$ hyperons
are combined to form candidates for
three decays of the charged $\Xi_c$,
\begin{eqnarray}
\label{XIPIPI}
\Xi_c^+  &  \to  &  \Xi^-\pi^+\pi^+ \\
\label{L0KAPIPI}
\Xi_c^+  &  \to  &  \Lambda K^-\pi^+\pi^+ \\
\label{PK0K0}
\Xi_c^+  &  \to  &   p K^0_S K^0_S 
\end{eqnarray} 
and four decays of the neutral state,
\begin{eqnarray} 
\label{XIPI}
\Xi_c^0  & \to  & \Xi^-\pi^+ \\
\label{L0KAPI}
\Xi_c^0  & \to  & \Lambda K^-\pi^+ \\ 
\label{L0K0}
\Xi_c^0  & \to  & \Lambda K^0_S \\ 
\label{PKKPI}
\Xi_c^0  & \to  & p K^- K^-\pi^+. 
\end{eqnarray} 
Combinatorial and $B\overline{B}$ backgrounds are suppressed by requiring 
that the momentum of the $\Xi_c$ candidate in the $e^+e^-$ 
center-of-mass system exceed 2.5~GeV/$c$.
The decay products are fitted to a common vertex, and a goodness-of-fit
criterion is applied:
for  decays~(\ref{L0KAPIPI}), (\ref{PK0K0}), (\ref{L0KAPI}) and~(\ref{L0K0}), 
which contain a $V^0$  ($\Lambda$ or $K^0_S$) in the final state,
we require $\chi^2/n.d.f. < 10$;
for the remaining decays, we require $\chi^2/n.d.f. < 50$.

%\begin{figure}[p]
%%\begin{figure}[hb]
%\setlength{\unitlength}{1mm}
%\begin{center}
%%\begin{picture}(140,60)(0,-35)
%%%\includegraphics[height=6.0cm,width=12cm]{./eps/fig1.eps}
%\epsfig{file=./eps/fig1.eps,width=12cm}
%\end{picture}
%\end{center}
%\caption{Feynman diagrams for the process $\Xi_c^0\to \Lambda K^0_S$:
% ({\bf a}) W exchange and ({\bf b})
%internal spectator.}
%\label{DIAGL0K0}
%%\end{figure}

\begin{figure}[htb]
\setlength{\unitlength}{1mm}
\begin{center}
\begin{picture}(130,85)(0,0)
\put(-1,30){\rotatebox{90}{\bf Events / (5~{\rm MeV}/$c^2$)}}
%\put(50,1){{\large\boldmath $m(\Xi^-\pi^+\pi^+)$}}
\put(50,1){{\large $m(\Xi\pi\pi)$ [GeV/$c^2$]}}
\includegraphics[height=8.5cm,width=12cm]{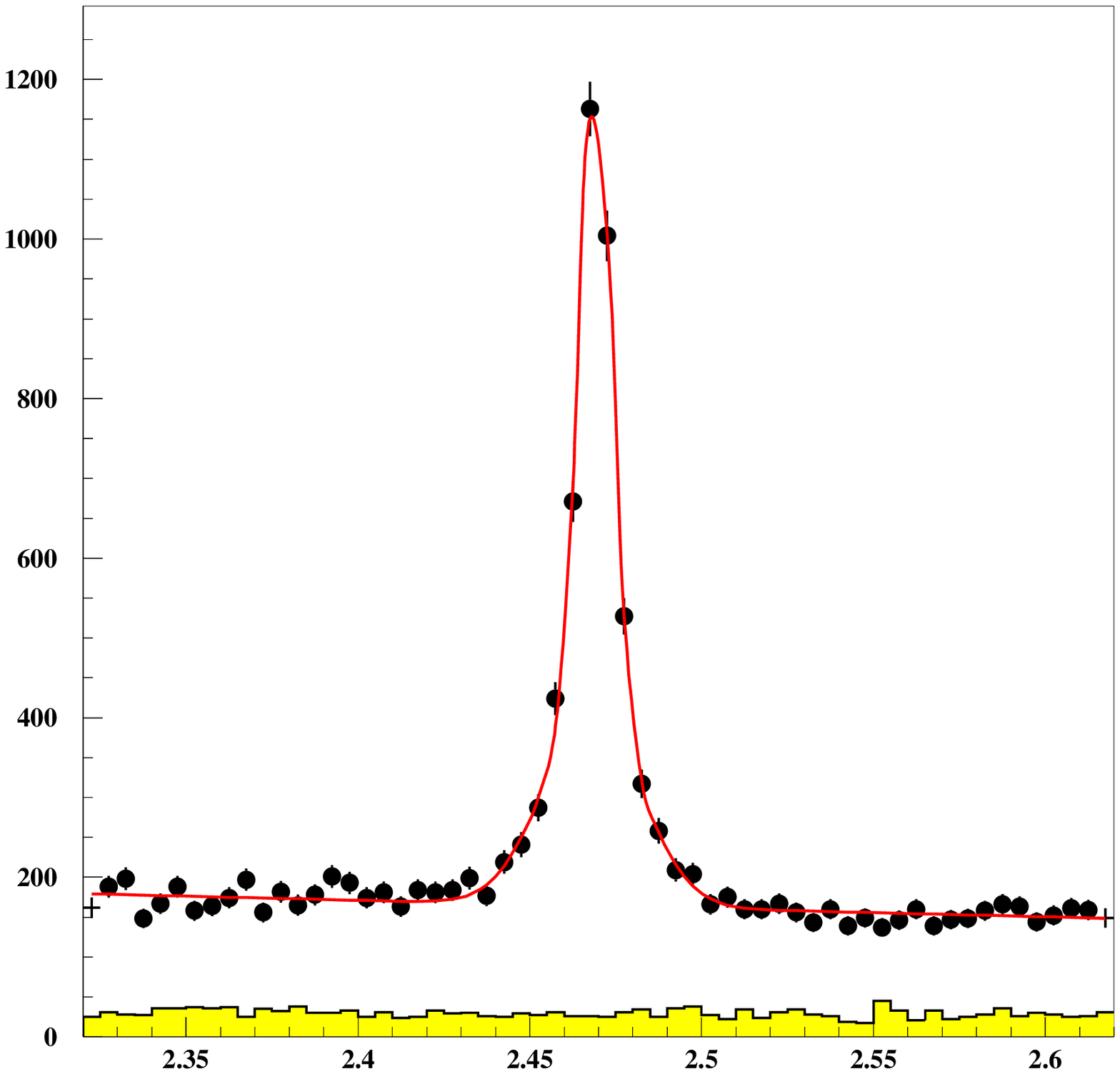}
\end{picture}
\end{center}
\caption{Invariant mass distribution of selected $\Xi^-(\to \Lambda\pi^-)\pi^+\pi^+$ combinations (points),
the fit described in the text (curve),
and wrong-sign combinations $(\overline{\Lambda} \pi^-)\pi^+\pi^+$ (shaded).}
\label{FIG_XIPIPI}
\end{figure}

\begin{figure}[p]
\setlength{\unitlength}{1mm}
\begin{center}
\begin{picture}(130,85)(0,0)
\put(-1,30){\rotatebox{90}{\bf Events / (2.5~{\rm MeV}/$c^2$)}}
%\put(50,1){{\large\boldmath $m(\Xi^-\pi^+\pi^+)$}}
\put(50,1){{\large $m(\Lambda K \pi\pi)$ [GeV/$c^2$]}}
\includegraphics[height=8.5cm,width=12cm]{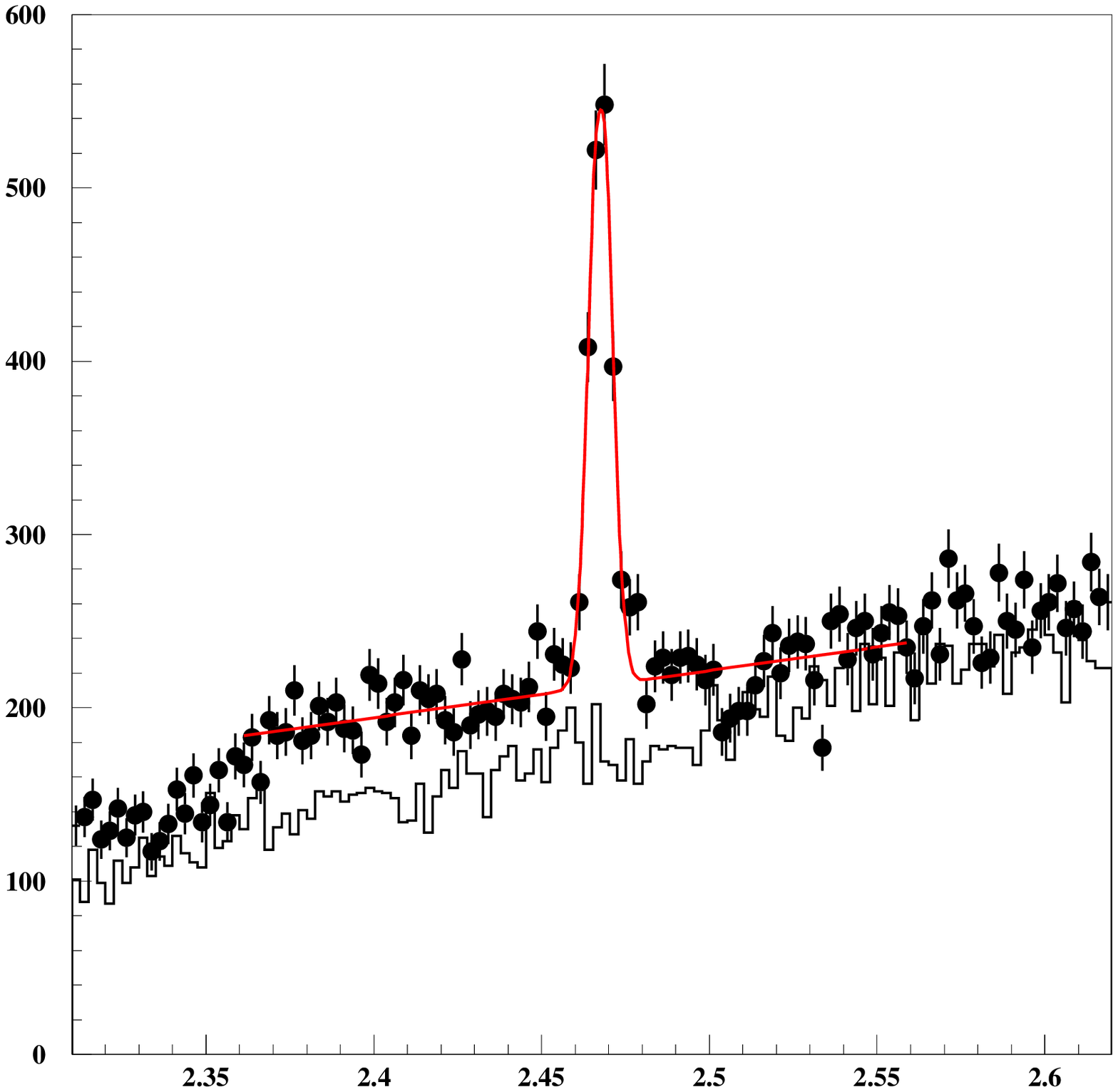}
\end{picture}
\end{center}
\caption{Invariant mass distribution of selected $\Lambda K^- \pi^+\pi^+$ combinations (points), 
the fit described in the text (curve), and wrong-sign combinations $\Lambda K^+ \pi^-\pi^-$ (histogram).}
\label{FIG_L0KAPIPI}
%\end{figure}
%
%\begin{figure}[p]
\setlength{\unitlength}{1mm}
\begin{center}
\begin{picture}(130,85)(0,0)
\put(-1,30){\rotatebox{90}{\bf Events / (2.5~{\rm MeV}/$c^2$)}}
%\put(50,1){{\large\boldmath $m(\Xi^-\pi^+)$}}
\put(50,1){{\large $m(p K^0_S K^0_S)$ [GeV/$c^2$]}}
\includegraphics[height=8.5cm,width=12cm]{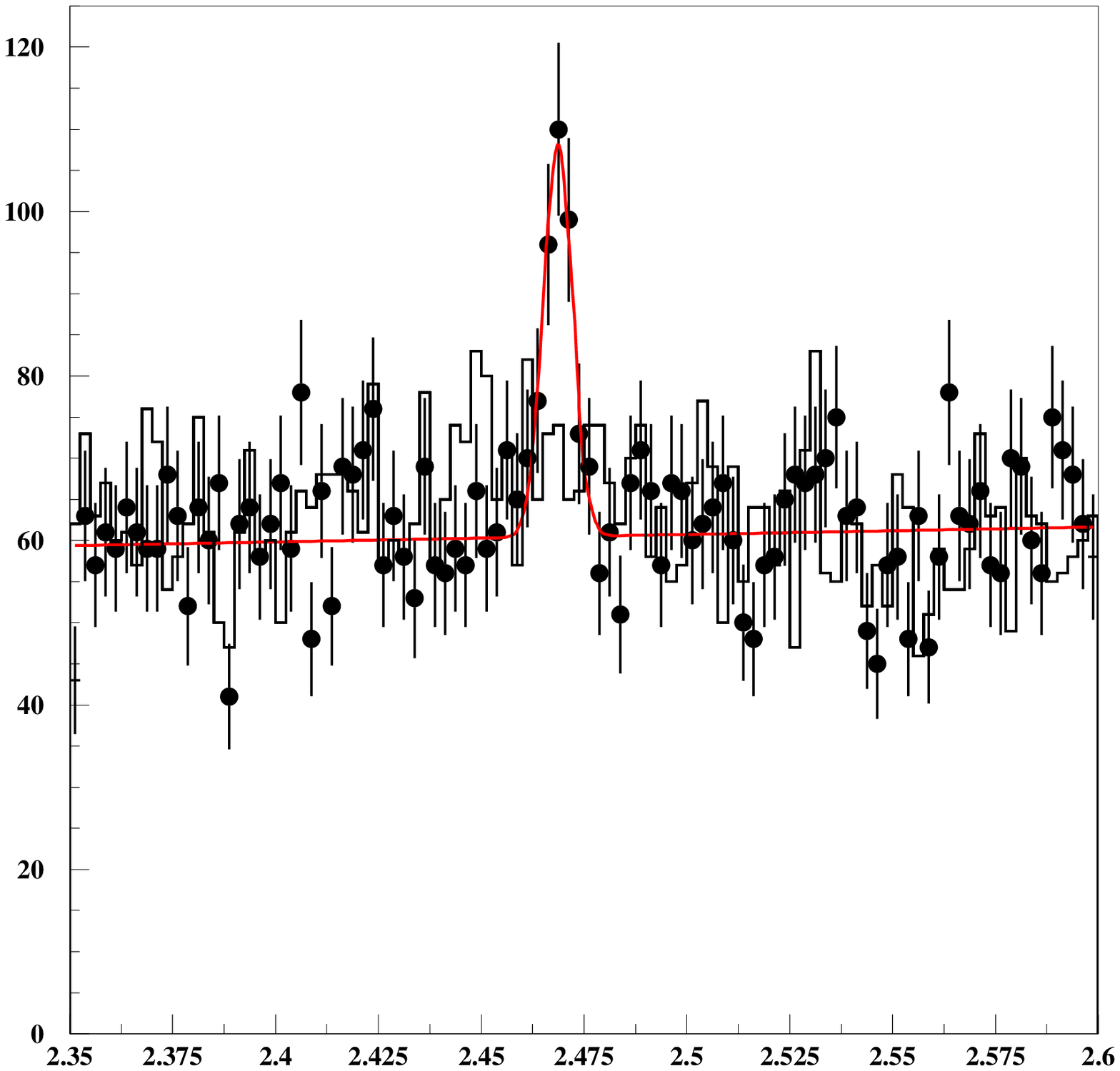}
\end{picture}
\end{center}
\caption{Invariant mass distribution of selected $p K^0_S K^0_S$ combinations (points),
the fit described in the text (curve), and $p K^0_S K^0_S$ combinations from
the $K^0_S\to\pi^+\pi^-$ mass sideband $11\,\mev/c^2 <
\left| m(\pi^+\pi^-) - 497.7\,\mev/c^2\right| < 17\,\mev/c^2$ (histogram).}
\label{FIG_PK0K0}
\end{figure}

\begin{figure}[p]
\setlength{\unitlength}{1mm}
\begin{center}
\begin{picture}(130,85)(0,0)
\put(-1,30){\rotatebox{90}{\bf Events / (5~{\rm MeV}/$c^2$)}}
%\put(50,1){{\large\boldmath $m(\Xi^-\pi^+)$}}
\put(50,1){{\large $m(\Xi\pi)$ [GeV/$c^2$]}}
\includegraphics[height=8.5cm,width=12cm]{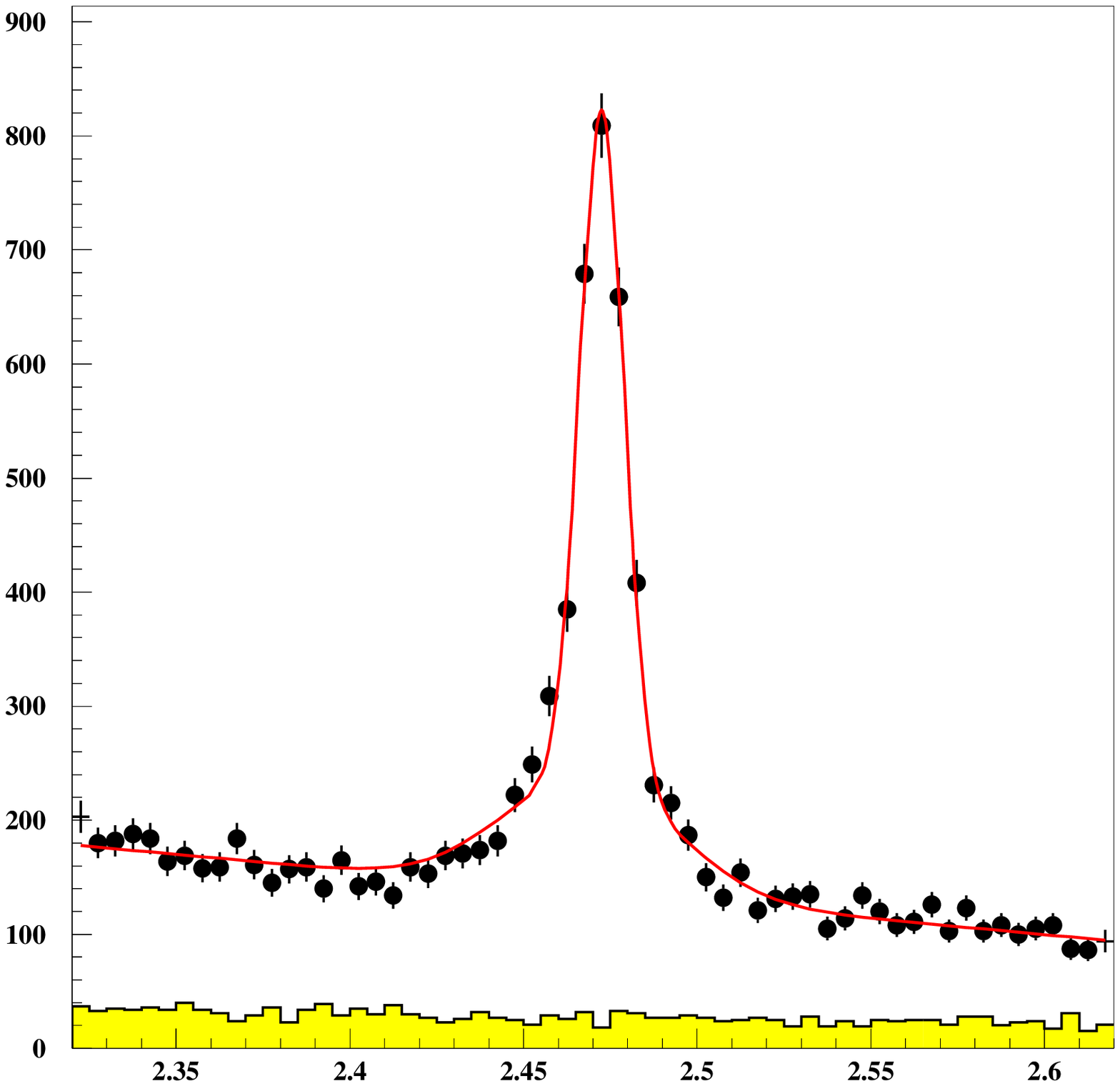}
\end{picture}
\end{center}
\caption{Invariant mass distribution of selected $\Xi^-(\to\Lambda\pi^-)\pi^+$ combinations (points), 
the fit described in the text (curve), and 
wrong-sign combinations $(\overline{\Lambda} \pi^-)\pi^+$ (shaded).}
\label{FIG_XIPI}
%\end{figure}
%
%\begin{figure}[p]
\setlength{\unitlength}{1mm}
\begin{center}
\begin{picture}(130,85)(0,0)
\put(-1,30){\rotatebox{90}{\bf Events / (5~{\rm MeV}/$c^2$)}}
%\put(50,1){{\large\boldmath $m(\Xi^-\pi^+)$}}
\put(50,1){{\large $m(\Lambda K \pi)$ [GeV/$c^2$]}}
\includegraphics[height=8.5cm,width=12cm]{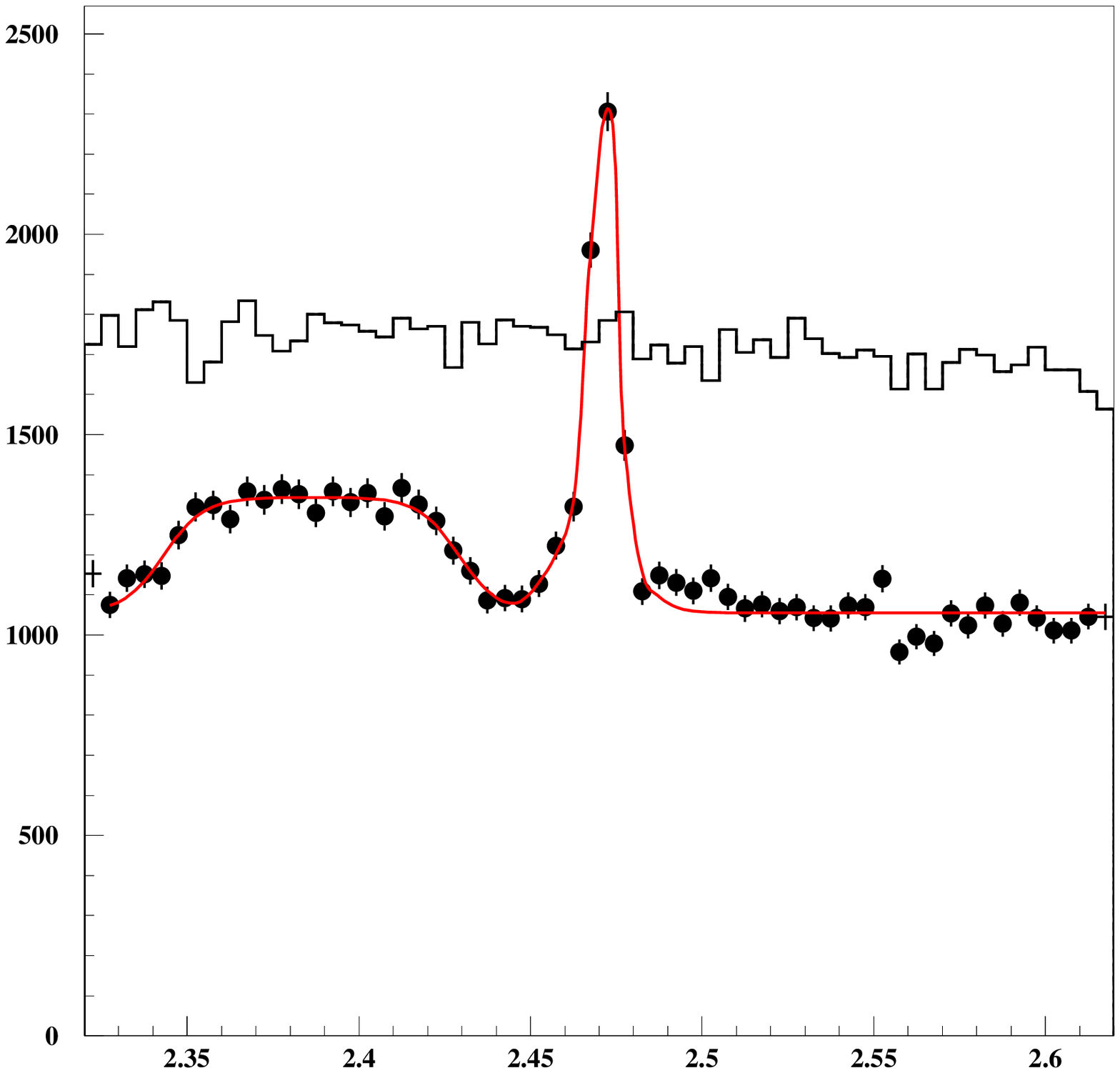}
\end{picture}
\end{center}
\caption{Invariant mass distribution of selected $\Lambda K^- \pi^+$ combinations (points), 
the fit described in the text (curve),
and wrong-sign combinations $\Lambda K^+ \pi^-$ (histogram).
The structure centered at 2.37 GeV/$c^2$ 
is discussed in the text.}
\label{FIG_L0KAPI}
\end{figure}

\begin{figure}[p]
\setlength{\unitlength}{1mm}
\begin{center}
\begin{picture}(130,85)(0,0)
\put(-1,30){\rotatebox{90}{\bf Events / (5~{\rm MeV}/$c^2$)}}
%\put(50,1){{\large\boldmath $m(\Xi^-\pi^+)$}}
\put(50,1){{\large $m(\Lambda K^0_S)$ [GeV/$c^2$]}}
\includegraphics[height=8.5cm,width=12cm]{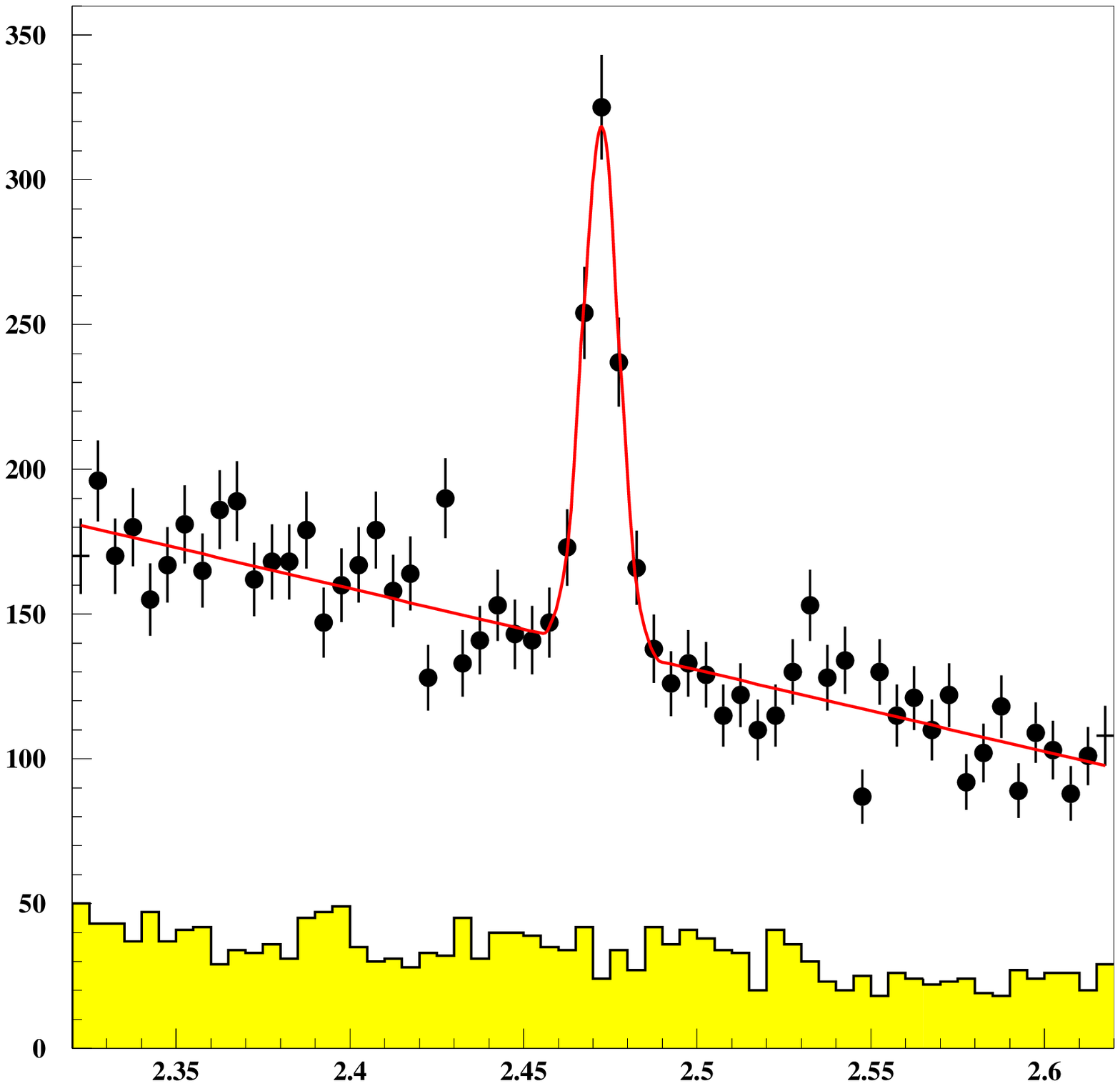}
\end{picture}
\end{center}
\caption{Invariant mass distribution of selected $\Lambda K^0_S$ combinations (points),
the fit described in the text (curve), and $\Lambda K^0_S$ combinations from  
the $K^0_S \to\pi^+\pi^-$ mass sideband $11\,\mev/c^2 < \left| m(\pi^+\pi^-) - 497.7\,\mev/c^2\right| < 17\,\mev/c^2$ (shaded).}
\label{FIG_L0K0}
%\end{figure}
%
%\begin{figure}[htb]
\setlength{\unitlength}{1mm}
\begin{center}
\begin{picture}(130,85)(0,0)
\put(-1,30){\rotatebox{90}{\bf Events / (2~{\rm MeV}/$c^2$)}}
%\put(50,1){{\large\boldmath $m(\Xi^-\pi^+)$}}
\put(50,1){{\large $m(p K K \pi)$ [GeV/$c^2$]}}
\includegraphics[height=8.5cm,width=12cm]{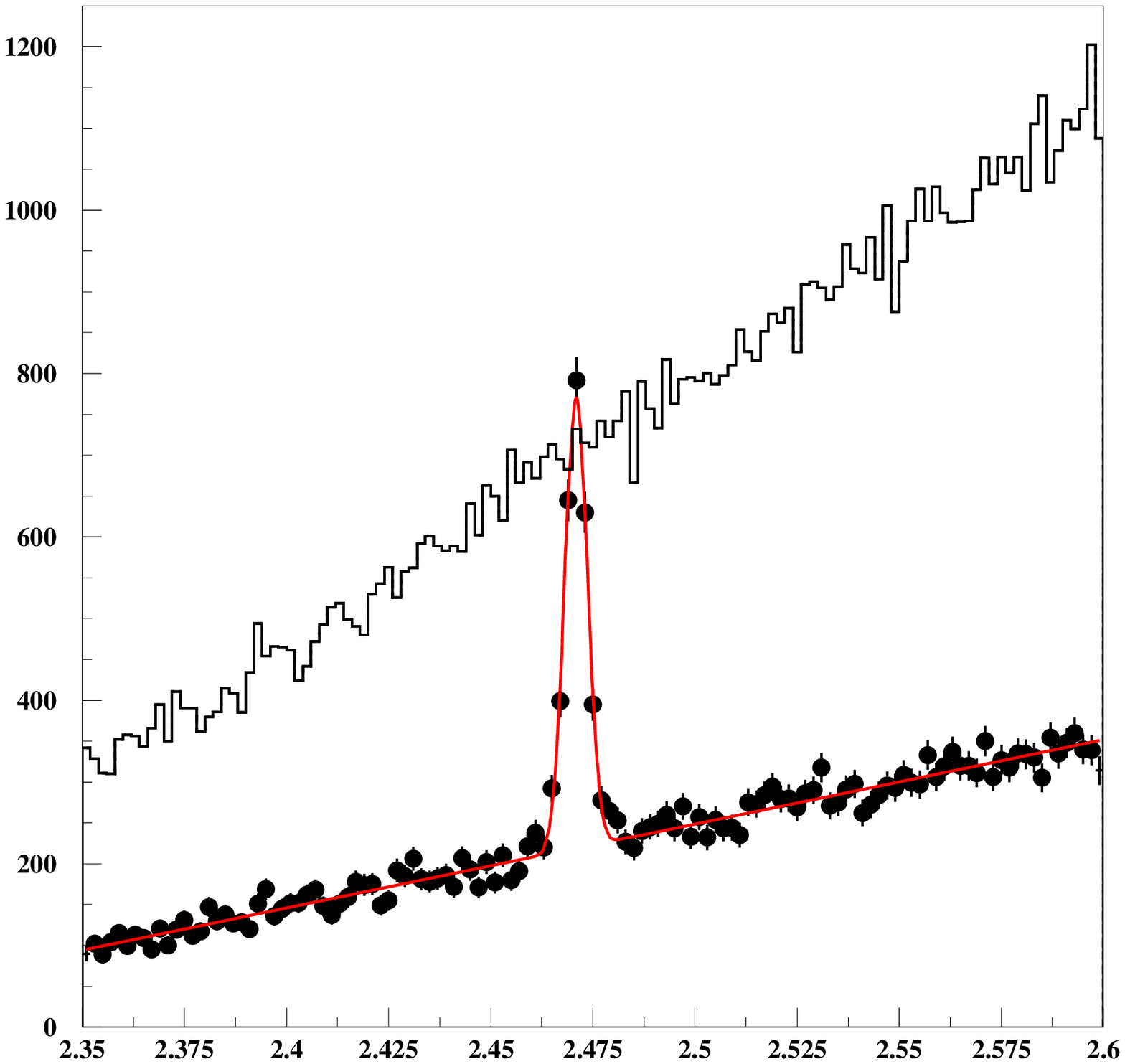}
\end{picture}
\end{center}
\caption{Invariant mass distribution of selected $p K^- K^-\pi^+$ combinations (points), 
the fit described in the text (curve),
and wrong-sign combinations $p K^- K^+\pi^-$ (histogram).}
\label{FIG_PKKPI}
\end{figure}

A clear~$\Xi_c$ baryon signal is observed in the invariant mass distributions
of each of the decays studied (Figs.~\ref{FIG_XIPIPI}--\ref{FIG_PKKPI}).
In particular, we observe the first evidence for the decay $\Xi_c^+ \to p K^0_S K^0_S$.
%(with significance of 7.5 standard deviations). 
For each decay mode, we extract the signal yield and the $\Xi_c$ mass and width
from a fit to the distribution.

For the decays
$\Xi_c^+\to \Xi^-\pi^+\pi^+$,
$\Xi_c^0\to \Xi^-\pi^+$ and
$\Xi_c^0\to \Lambda K^- \pi^+$, 
we use a double Gaussian for the signal
(the second Gaussian is required to account for the tails in the signal shape)
and a linear background function.
In each case, the means of both Gaussians coincide within the errors of the fits.
The broad enhancement in the $M(\Lambda K\pi)$ distribution
(Fig.~\ref{FIG_L0KAPI}), below the $\Xi_c$ mass,
is assumed to be due to $\Xi(2370) \to \Lambda K\pi$ decays with 
an admixture of a kinematic reflection;
it is represented with  the function
$f(x) = 1/(1+\exp( (\left| x-\overline{x} \right| - \sigma_x)/z  ) )$~\cite{BELL},
where $x$ is the $\Lambda K \pi$ invariant mass and
$\overline{x}$, $\sigma_x$ and $z$ are parameters allowed to float in the fit.
The mass and width of this structure are 
compatible with the parameters of the~$\Xi(2370)$ resonance~\cite{PDG}.

For the decays
$\Xi_c^+\to \Lambda K^-\pi^+\pi^+$, 
$\Xi_c^+\to  p K^0_S K^0_S$,
$\Xi_c^0\to \Lambda K^0_S$
and $\Xi_c^0\to p K^- K^- \pi^+$,
we use a single Gaussian
for the signal and a linear background function.
In each case the $\Xi_c$ width is found to be compatible with
the value from Monte Carlo simulation. 

Figures~\ref{FIG_XIPIPI},~\ref{FIG_L0KAPIPI},~\ref{FIG_XIPI},~\ref{FIG_L0KAPI} and~\ref{FIG_PKKPI}
also show the distribution of ``wrong-sign'' combinations, 
for which charge-conjugate states are used for certain particles. 
Figures~\ref{FIG_PK0K0} and~\ref{FIG_L0K0} include the invariant mass spectrum
for $\Xi_c$ candidates taken from $K^0_S\to\pi^+\pi^-$ mass sidebands. 
These distributions are structureless and provide a cross-check for the
shape of the combinatorial background in the $\Xi_c$ sample.

The fit results are summarized in Table~\ref{RESXC}.
For each mode where a double Gaussian parametrization is used,
the~$\Xi_c$ mass is taken as the average of the means of the two Gaussians,
weighted by their yields. 

\begin{table}[htb]
\begin{center}
\caption{Signal yields, fitted $\Xi_c$ masses, and reconstruction efficiencies
for the $\Xi_c$ decay analyses described in the text.}
\vspace*{0.5ex}
\begin{tabular}{lrcr}
\hline
%Decay  & \# of  events	&  mass		& \multicolumn{1}{c}{Efficiency}\\ 
%       & \multicolumn{1}{c}{$N_{\rm sig}$}
%			&  [MeV/$c^2$]	& \multicolumn{1}{c}{[\%]}	\\
Decay mode			&  \# of  events	& mass [MeV/$c^2$]		& \multicolumn{1}{c}{Efficiency [\%]}\\ 
\hline
$\Xi_c^+\to \Xi^-\pi^+\pi^+$	&$3605\pm 279$		&  $2468.6 \pm 0.4\pm 0.5$	& $4.55\pm 0.07$	\\
$\Xi_c^+\to \Lambda K^- \pi^+\pi^+$
				&$1177\pm\phantom{2}55$ &  $2467.6 \pm 0.2 \pm 0.5$	& $4.70\pm 0.10$	\\
$\Xi_c^+\to p K^0_S K^0_S$	& $168\pm\phantom{2}27$ &  $2468.6 \pm 0.7 \pm 0.9$	& $2.45\pm 0.06$	\\
\hline
$\Xi_c^0\to \Xi^-\pi^+$		&$2979\pm 211$		&  $2471.3 \pm 0.5 \pm 0.8$ 	& $7.13\pm 0.14$	\\
$\Xi_c^0\to \Lambda K^-\pi^+$	&$3268\pm 276$		&  $2470.0 \pm 0.6 \pm 0.7$	& $7.31\pm 0.11$	\\
$\Xi_c^0\to \Lambda K^0_S$	& $465\pm\phantom{2}37$ &  $2472.2 \pm 0.5 \pm 0.5$	& $5.36\pm 0.12$	\\
$\Xi_c^0\to p K^- K^- \pi^+$	&$1908\pm\phantom{2}62$ &  $2470.9 \pm 0.1 \pm 0.2$	& $14.00\pm 0.20$	\\
\hline
\end{tabular}
\label{RESXC}
\end{center}
\end{table}

%%%%%%%%%%%%%%%%%%%%%%%%%%%%%%%%%%%%%%%%%%%%%%%%%%%%%%%%%%%%%%%%%%%%%%%%%%%%%%

\section{$\Xi_c$ mass determination}
\label{MASS}

%%%%%%%%%%%%%%%%%%%%%%%%%%%%%%%%%%%%%%%%%%%%%%%%%%%%%%%%%%%%%%%%%%%%%%%%%%%%%%

The average masses of the~$\Xi_c^0$ and~$\Xi_c^+$ are determined from
the values in Table~\ref{RESXC}
using the PDG unconstrained averaging algorithm~\cite[Introduction, p.14--15]{PDG}: 
\begin{align} 
m_{\Xi_c^+} & = (2468.1\pm 0.4\,{\rm (stat. \oplus syst.)}^{+1.4}_{-0.2})\,\mev/c^2 && ({\rm PDG:} & (2466.3\pm 1.4)\,\mev/c^2)\phantom{;} 
\label{MASSXCP}	\\
m_{\Xi_c^0} & = (2471.0\pm 0.3\,{\rm (stat. \oplus syst.)}^{+1.4}_{-0.2})\,\mev/c^2 && ({\rm PDG:} & (2471.8\pm 1.4)\,\mev/c^2); 
\label{MASSXC0}
\end{align} 
the first error is the combined statistical and systematic uncertainty,
and the second is the uncertainty due to possible biasses in the mass scale
(discussed below).
We therefore find the $\Xi_c^0 - \Xi_c^+$ mass splitting to be
\begin{equation} 
m_{\Xi_c^0} - m_{\Xi_c^+} = (2.9\pm 0.5)\,\mev/c^2~~~~({\rm PDG:}~~(5.5\pm 1.4)\,\mev/c^2). 
\label{MASSSPL}
\end{equation} 
The systematic uncertainty in the mass determination is evaluated as follows.
For each mode we vary the order of the polynomial describing the background
and the mass range covered by the fit, yielding changes in the fitted
mass between 0.1 and $0.5\,\mev/c^2$, depending on the decay.
To model imperfect understanding of the signal resolution, we perform 
fits using signal widths fixed from Monte Carlo,
and compare with values where the widths are floated: the mass 
changes by 0.1--$0.5\,\mev/c^2$.
Varying the selection criteria, we find a corresponding uncertainty
of 0.2--$0.8\,\mev/c^2$.
To study the possible dependence of the $\Xi_c$ mass on the momentum and decay length
of the $V^0$'s, the mass is estimated in bins of these variables,
and an uncertainty of $0.4\,\mev/c^2$ is assigned.
The total systematic uncertainty is obtained by adding the individual contributions in quadrature. 
%(cf. Table~\ref{SYSTMASS}).

The possible bias in the overall mass scale is estimated using two approaches. 
First we reconstruct the following decays, kinematically
similar to the ones under study:
 $\Lambda_c^+\to p K^- \pi^+$, $\Lambda_c^+\to \Lambda\pi^+\pi^+\pi^-$, $D^0\to K^0_S K^0_S$ and
$D^+\to K^+K^0_S K^0_S$.
Comparison of the fitted masses of parent particles with world-average 
values~\cite{PDG} yields a maximum mass shift of $+1.4\,\mev/c^2$.
Second, using Monte Carlo samples, generated and reconstructed masses 
are compared for each of the decays~(\ref{XIPIPI})--(\ref{PKKPI}),
yielding a maximum shift of $\pm 0.2\,\mev/c^2$.
As a result, $^{+1.4}_{-0.2}\,\mev/c^2$ is assigned as a measure
of uncertainty in the overall mass scale. 
Any such shift is assumed to  cancel in the mass splitting  $m_{\Xi_c^0}-m_{\Xi_c^+}$ (Eq.~\ref{MASSSPL}).

%\begin{table}[htb]
%\begin{center}
%\caption{Systematic uncertainties of the mass determination (in MeV/$c^2$).}
%\begin{tabular}{lccccc}
%\hline
%Decay  &  Background  & Signal  &  Selection  & $V^0$ momentum       &  Total         \\ 
%       &  shape       & width   &  cuts       & and flight-distance  &  uncertainty   \\
%\hline
%$\Xi_c^+\to \Xi^-\pi^+\pi^+$           &  0.1 & 0.1  & 0.2 &  0.4 & 0.5 \\
%$\Xi_c^+\to \Lambda K^- \pi^+\pi^+$  &  0.1 & 0.1  & 0.2 &  0.4 & 0.5 \\
%$\Xi_c^+\to p K^0_S K^0_S$             &  0.1 & 0.1  & 0.8 &  0.4 & 0.9 \\
%$\Xi_c^0\to \Xi^-\pi^+$                &  0.3 & 0.5  & 0.2 &  0.4 & 0.8 \\
%$\Xi_c^0\to \Lambda K^-\pi^+$        &  0.5 & 0.1  & 0.3 &  0.4 & 0.7 \\
%$\Xi_c^0\to \Lambda K^0_S$           &  0.1 & 0.2  & 0.1 &  0.4 & 0.5 \\
%$\Xi_c^0\to p K^- K^- \pi^+$           &  0.1 & 0.1  & 0.1 &   ~0 & 0.2 \\
%\hline
%\end{tabular}
%\label{SYSTMASS}
%\end{center}
%\end{table}

%%%%%%%%%%%%%%%%%%%%%%%%%%%%%%%%%%%%%%%%%%%%%%%%%%%%%%%%%%%%%%%%%%%%%%%%%%%%%%

\section{$\Xi_c$ branching ratios}
\label{BRANCHING}

%%%%%%%%%%%%%%%%%%%%%%%%%%%%%%%%%%%%%%%%%%%%%%%%%%%%%%%%%%%%%%%%%%%%%%%%%%%%%%

Branching ratios are evaluated by comparing signal yields for the relevant
decays, correcting for reconstruction efficiencies as determined from Monte
Carlo (Table~\ref{RESXC});
%(see Table~\ref{EFFIC}).
branching fractions for the intermediate decays
 $\Lambda\to p \pi^-$ and  $K^0_S\to\pi^+\pi^-$
are taken into account.
%(cf. Table~\ref{EFFIC}). 
The modes $\Xi_c^+\to \Xi^-\pi^+\pi^+$ and $\Xi_c^0\to\Xi^-\pi^+$ are used as 
references, yielding

\begin{align} 
\label{BRL0KAPIPI}
\frac{\Gamma (\Xi_c^+\to \Lambda K^-\pi^+\pi^+)}{\Gamma (\Xi_c^+\to\Xi^-\pi^+\pi^+)}            & =    0.32\pm 0.03 \pm 0.02  \\[0.8ex]
\label{BRPK0K0}
\frac{\Gamma (\Xi_c^+\to p  K^0_S K^0_S)}{\Gamma (\Xi_c^+\to\Xi^-\pi^+\pi^+)}    & =    0.087\pm 0.016 \pm 0.014  \\[0.8ex]
\label{BRL0KAPI}
\frac{\Gamma (\Xi_c^0\to \Lambda K^-\pi^+)}{\Gamma (\Xi_c^0\to\Xi^-\pi^+)}                  & =    1.07\pm 0.12 \pm 0.07  \\[0.8ex]
\label{BRL0K0}
\frac{\Gamma (\Xi_c^0\to \Lambda K^0_S)}{\Gamma (\Xi_c^0\to\Xi^-\pi^+)}           & =    0.21\pm 0.02 \pm 0.02  \\[0.8ex]
\label{BRPKKPI}
\frac{\Gamma (\Xi_c^0\to  p K^- K^-\pi^+)}{\Gamma (\Xi_c^0\to\Xi^-\pi^+)}                      & =    0.33\pm 0.03 \pm 0.03. 
\end{align}

%\begin{table}[h]
%\begin{center}
%\caption{Reconstruction efficiencies ($\epsilon$) for the $\Xi_c$ baryon decay modes and
% efficiency corrected numbers of signal events $N_{sig}/\epsilon$ 
%(cf. Table~\ref{RESXC} for $N_{sig}$).}
%\begin{tabular}{lrr}
%\hline
%Decay  &  Efficiency             &  $N_{sig}/\epsilon$~~~ \\
%       &  $\epsilon$ (\% )       &                        \\
%\hline
%$\Xi_c^+\to \Xi^-\pi^+\pi^+$            &  $2.9\pm 0.1$ & $ 125124\pm 9678 $   \\
%$\Xi_c^+\to \Lambda K^-\pi^+\pi^+$    &  $2.4\pm 0.1$ & $  48321\pm 2302 $   \\
%$\Xi_c^+\to p K^0_S K^0_S$              &  $1.5\pm 0.1$ & $  11052\pm 1790 $   \\
%\hline
%$\Xi_c^0\to \Xi^-\pi^+$                 &  $4.6\pm 0.1$ & $  65456\pm 4646 $   \\
%$\Xi_c^0\to \Lambda K^- \pi^+$        &  $4.7\pm 0.1$ & $  69001\pm 5856 $   \\
%$\Xi_c^0\to \Lambda K^0_S$            &  $2.3\pm 0.1$ & $  19802\pm 1558 $   \\
%$\Xi_c^0\to p K^- K^- \pi^+$            & $14.0\pm 0.2$ & $  13632\pm~~443 $   \\
%\hline
%\end{tabular}
%\label{EFFIC}
%\end{center}
%\end{table}

\begin{table}[h]
\begin{center}
\caption{Systematic uncertainties on the signal yields;
errors are given in percent (\%).}
\vspace*{0.5ex}
\begin{tabular}{lccccc}
\hline
%Decay  &  Background  & Signal  & Monte Carlo  &  Fragmentation & Total         \\ 
%       &  shape       & width   & statistics   &  function      & uncertainty   \\
Decay mode  &  Bkgd shape & Signal width & MC stats  &  Fragmentation & Total         \\ 
\hline
$\Xi_c^+\to \Xi^-\pi^+\pi^+$           &  1.2 & 0.5  & 1.1 & 1.0 & 2.0 \\
$\Xi_c^+\to \Lambda K^- \pi^+\pi^+$  &  1.4 & 3.4  & 2.2 & 1.0 & 4.4 \\
$\Xi_c^+\to p K^0_S K^0_S$             &  1.4 & 3.0  & 2.5 & 1.0 & 4.3 \\
$\Xi_c^0\to \Xi^-\pi^+$                &  1.6 & 4.2  & 2.1 & 1.0 & 5.1 \\
$\Xi_c^0\to \Lambda K^-\pi^+$        &  0.8 & 1.3  & 1.2 & 1.0 & 2.2 \\
$\Xi_c^0\to \Lambda K^0$             &  1.3 & 3.2  & 2.5 & 1.0 & 4.4 \\
$\Xi_c^0\to p K^- K^- \pi^+$           &  0.3 & 1.5  & 1.1 & 1.0 & 2.1 \\
\hline
\end{tabular}
\label{SYSTBRFR}
\end{center}
\end{table}

\begin{table}[h]
\begin{center}
\caption{Systematic uncertainties on the branching ratios;
errors are given in percent (\%).}
\vspace*{0.5ex}
\begin{tabular}{lccccc}
\hline
% Branching   &  Syst. err. of & Syst. err. of & $V^0$             & Hadron   &  Total         \\ 
% ratio       &    numerator   & denominator   & recon.             & iden.  &  uncertainty   \\
Branching ratio &  Numerator & Denominator & $V^0$ recon.         & Hadron ID  &  Total         \\ 
\hline
$\frac{\Gamma(\Xi_c^+\to \Lambda K^- \pi^+\pi^+)}{\Gamma(\Xi_c^+\to \Xi^-\pi^+\pi^+)}$ &  4.4 & 2.0  & ~0.0  & 3.0 & ~5.7 \\
$\frac{\Gamma(\Xi_c^+\to p K^0_S K^0_S)}{\Gamma(\Xi_c^+\to \Xi^-\pi^+\pi^+)}$            &  4.3 & 2.0  & 15.0  & 0.0 & 15.7 \\
$\frac{\Gamma(\Xi_c^0\to \Lambda K^-\pi^+)}{\Gamma(\Xi_c^0\to \Xi^-\pi^+)}$            &  2.2 & 5.1  & ~0.0  & 3.0 & ~6.3 \\
$\frac{\Gamma(\Xi_c^0\to \Lambda K^0_S)}{\Gamma(\Xi_c^0\to \Xi^-\pi^+)}$               &  4.4 & 5.1  & ~5.0  & 0.0 & ~8.4 \\
$\frac{\Gamma(\Xi_c^0\to p K^- K^- \pi^+)}{\Gamma(\Xi_c^0\to \Xi^-\pi^+)}$               &  2.1 & 5.1  & ~5.0  & 6.0 & ~9.6 \\
\hline
\end{tabular}
\label{SYSTBRRT}
\end{center}
\end{table}

The following sources of systematic error on the  efficiency corrected signal
yields are considered:
uncertainties in the background
shape and signal width  (evaluated as described in the previous section),
uncertainty due to the limited statistics of Monte Carlo samples used to 
determine efficiencies,
and the uncertainty due to charm fragmentation~\cite{CHARM}.
The latter contribution
is estimated using the deviations between the data and Monte Carlo simulation
for samples containing $D^*$ mesons, modelled by 
the fragmentation function of Peterson et al.~\cite{PETERSON}.
The resulting uncertainties are summarized in Table~\ref{SYSTBRFR};
the totals are obtained by adding the individual contributions in quadrature.

When determining the branching ratios, uncertainties due to reconstruction of $V^0$'s and particle identification
of hadrons are taken into account. Based on a comparison
of yields for the decays 
$D^+\to  K^0_S\pi^+$ and $D^+\to K^-\pi^+\pi^+$ in data and Monte Carlo,
%~\cite{NAKAO},
 the uncertainty on the efficiency of $K^0_S$ finding
was estimated to be 5.0\%. The same value was assigned
for~$\Lambda$ finding;
in the case~(\ref{BRPK0K0}) we conservatively assume that any such error is
anti-correlated with that due to $K^0_S$ finding,
and add the uncertainties linearly.
Uncertainties on particle identification efficiency are taken to be
1\% for each pion and
%(vetoed only),
2\% for each kaon;
for the cases~(\ref{BRL0KAPIPI}) and~(\ref{BRL0KAPI}),
we add these uncertainties linearly.
Any error in proton identification efficiencies is assumed to cancel
in the ratios~(\ref{BRL0KAPIPI})--(\ref{BRPKKPI}).
For each branching ratio, the systematic uncertainties from these sources,
and from the uncertainties in the yields of the two decay modes
(``numerator'' and ``denominator''; see Table~\ref{SYSTBRFR}),
are summarized in Table~\ref{SYSTBRRT}.
The total uncertainty is obtained by combining each term in quadrature.

The branching ratio given in Eq.~(\ref{BRL0KAPIPI}) is consistent with the  
recent FOCUS measurement $0.28 \pm 0.06 \pm 0.06$~\cite{FOCUS03},
and somewhat lower than the previous CLEO result
$0.58\pm 0.16\pm 0.07$~\cite{CLEO96a};
the ratio~(\ref{BRPKKPI}) is compatible with
the CLEO result $0.35\pm 0.06\pm 0.03$~\cite{CLEO03}.
The three remaining branching ratios (\ref{BRPK0K0},~\ref{BRL0KAPI} and~\ref{BRL0K0}) are measured for the first time.
The branching ratio for the decay $\Xi_c^0\to \Lambda K^0_S$ is in agreement with the existing theoretical 
predictions~\cite{SYMQUA1,SYMQUA2,ZENCZYK,SHARMA}.
 This measurement is in fact more precise than the current range of theoretical predictions
and hence can potentially significantly constrain the above models.

\subsection{Resonant substructures in the decay $\Xi_c^0\to p K^- K^- \pi^+ $}
\label{SUBSTRUCTURE}

\begin{figure}[p]
\setlength{\unitlength}{1mm}
\begin{center}
\begin{picture}(130,85)(0,0)
\put(-1,30){\rotatebox{90}{\bf Events / (10~MeV/$c^2$)}}
%\put(50,1){{\large\boldmath $m(\Xi^-\pi^+)$}}
\put(15,71){{\large {\bf (a)}}}
\put(15,33){{\large {\bf (b)}}}
\put(50,1){{\large $m(K \pi)$ [GeV/$c^2$]}}
\includegraphics[height=8.5cm,width=12cm]{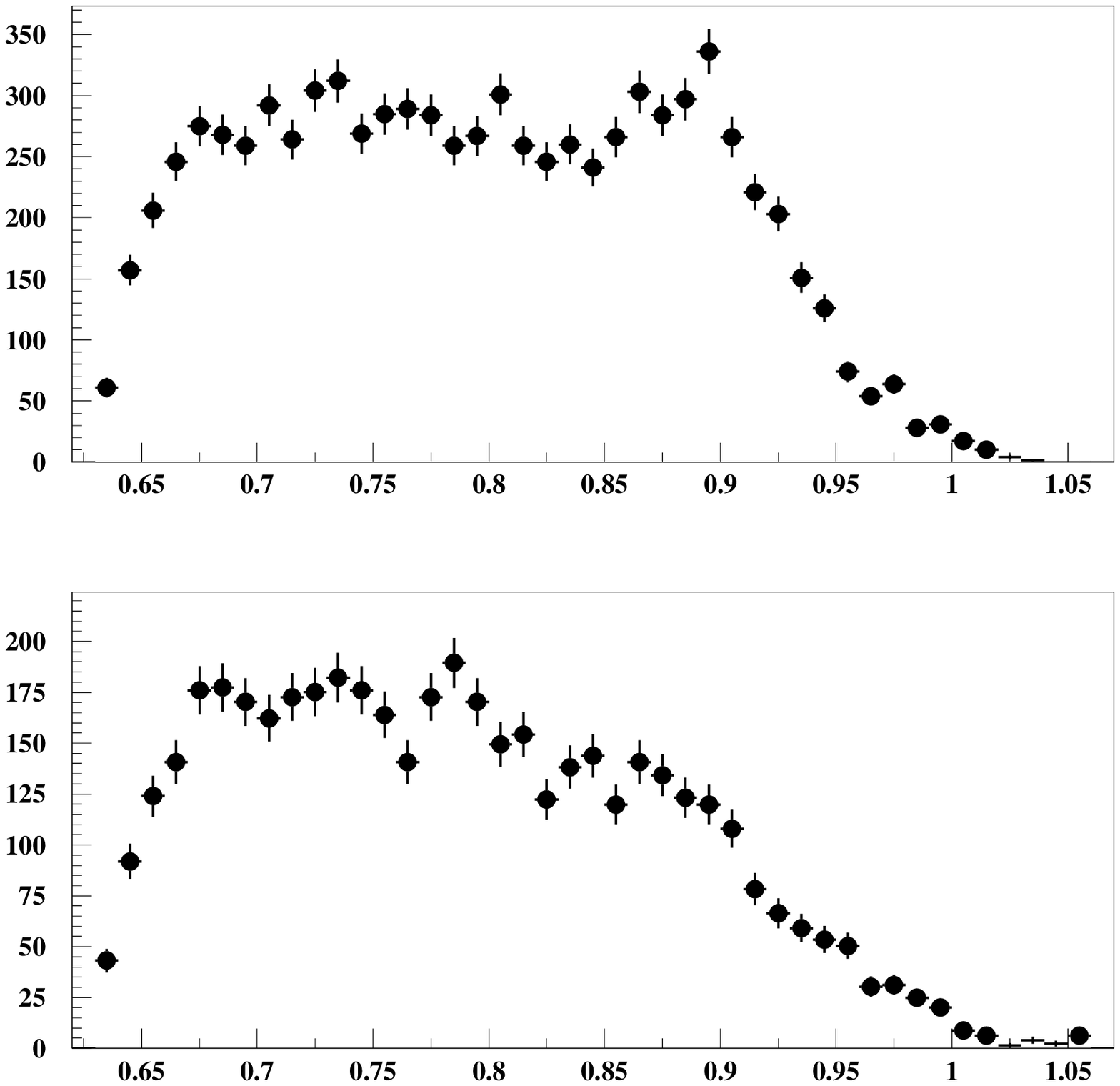}
\end{picture}
\end{center}
\caption{$\Xi_c^0 \to p K^- K^- \pi^+$: invariant mass distribution
of $K^-\pi^+$ pairs from (a) the $\Xi_c^0$ peak and (b) the $\Xi_c^0$ mass
sidebands, normalized to the background below the $\Xi_c^0$ peak.}
\label{FIG_PKKPIRES_0}
%\end{figure}
%\begin{figure}[th]
\setlength{\unitlength}{1mm}
\begin{center}
\begin{picture}(130,85)(0,0)
\put(-1,30){\rotatebox{90}{\bf Events / (10~MeV/$c^2$)}}
%\put(50,1){{\large\boldmath $m(\Xi^-\pi^+)$}}
\put(50,1){{\large $m(K \pi)$ [GeV/$c^2$]}}
\includegraphics[height=8.5cm,width=12cm]{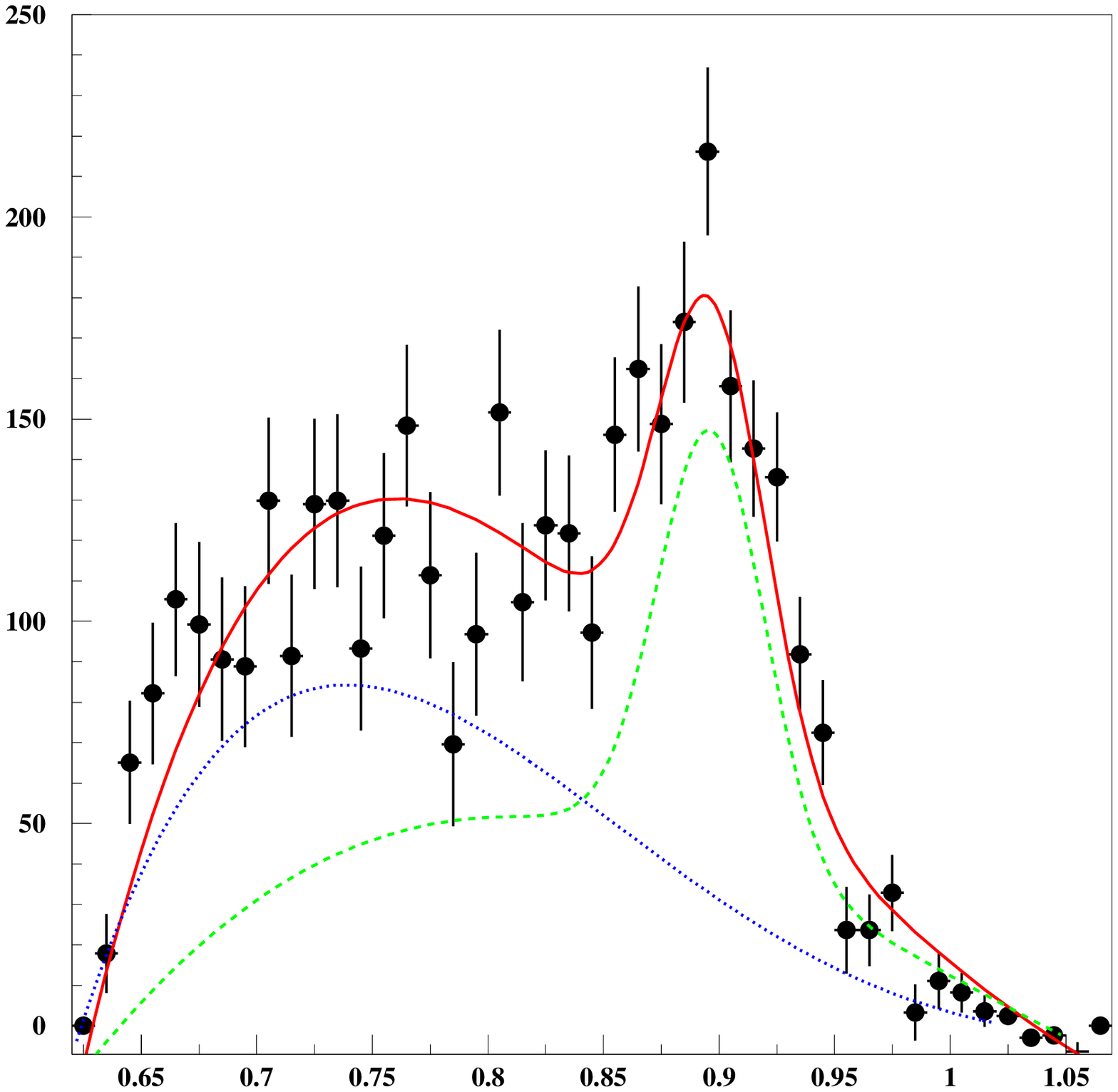}
\end{picture}
\end{center}
\caption{$\Xi_c^0 \to p K^- K^- \pi^+$: background subtracted $K^-\pi^+$ 
invariant mass distribution (points),
the fit described in the text (solid curve),
the $\Xi_c^0\to p K^- \ksbarz$ component (dashed)
and the non-resonant $\Xi_c^0\to p K^- K^- \pi^+$ contribution (dotted).
The background is modelled using $\Xi_c^0$ mass sidebands;
see Fig.~\ref{FIG_PKKPIRES_0}. 
}
\label{FIG_PKKPIRES}
\end{figure}

In the $p K^- K^- \pi^+ $ final state, a search for the intermediate 
resonance \ksbarz\ is performed by examining the kaon-pion invariant mass distribution
for each of the two kaon candidates. This distribution
is formed for combinations $p K K \pi$ 
within three standard deviations  of the $\Xi_c^0$ mass peak,  
(2.462--2.482)~GeV/$c^2$ (Fig.~\ref{FIG_PKKPIRES_0}(a)),
and for combinations in the $\Xi_c^0$ mass sidebands
(2.427--2.44)~GeV/$c^2$ and (2.50--2.513)~GeV/$c^2$ 
(Fig.~\ref{FIG_PKKPIRES_0}(b)).
Fig.~\ref{FIG_PKKPIRES} shows the sideband-subtracted $K\pi$ invariant mass
spectrum together with a fit
to two components corresponding to resonant $\Xi_c^0\to p K^- \ksbarz$
and non-resonant $\Xi_c^0\to p K^- K^- \pi^+ $ decays respectively. The shapes of both spectra are determined from
Monte Carlo simulation. The decay $\Xi_c^0\to p K^- \ksbarz$ is generated according to a 3-body phase space distribution, and is well-described by 
a Gaussian and a fourth order polynomial;
the non-resonant contribution
is parametrized by a fourth order polynomial. In the fit, the only free parameter is the
 fraction of the resonant component. The fit yields a resonant fraction of 
$0.51\pm 0.03\pm 0.01$, where the systematic error is estimated by varying the parametrization
of the two components. The resonant fraction was also recently measured by the CLEO
experiment to be $0.39\pm 0.06$ (statistical error only).
No statistically significant signal for the two-body decay
 $\Xi_c^0\to \Lambda(1520) \ksbarz$, with the subsequent decays
$\Lambda(1520)\to p K^-$ and $\ksbarz\to K^-\pi^+$, 
is observed.

%%%%%%%%%%%%%%%%%%%%%%%%%%%%%%%%%%%%%%%%%%%%%%%%%%%%%%%%%%%%%%%%%%%%%%%%%%%%%%

\section{Conclusions}

%%%%%%%%%%%%%%%%%%%%%%%%%%%%%%%%%%%%%%%%%%%%%%%%%%%%%%%%%%%%%%%%%%%%%%%%%%%%%%
Seven exclusive decays of the $\Xi_c$ baryon are observed using data collected by the Belle
experiment. The masses of charged and neutral states are determined to be 
$2468.1\pm 0.4^{+1.4}_{-0.2}\,\mev/c^2$ and
$2471.0\pm 0.3^{+1.4}_{-0.2}\,\mev/c^2$, respectively,
and the mass splitting is measured to be
$m_{\Xi_c^0} - m_{\Xi_c^+} = (2.9\pm 0.5)\,\mev/c^2$.
Branching ratios relative to the modes
$\Xi_c^+\to \Xi^-\pi^+\pi^+$ and $\Xi_c^0\to \Xi^-\pi^+$ have also been determined.
The branching ratios
$\Gamma (\Xi_c^+\to \Lambda K\pi\pi)/\Gamma (\Xi_c^+\to\Xi\pi\pi)  = 0.32\pm 0.03 \pm 0.02$ and  
$\Gamma (\Xi_c^0\to  p KK\pi)/\Gamma (\Xi_c^0\to\Xi\pi)            = 0.33\pm 0.03 \pm 0.03$ 
confirm, with improved precision, previous results of the FOCUS~\cite{FOCUS03} and CLEO~\cite{CLEO96a,CLEO03} experiments.
The branching ratios 
$\Gamma (\Xi_c^+\to p  K^0_S K^0_S)/\Gamma (\Xi_c^+\to\Xi\pi\pi)     = 0.087\pm 0.016 \pm 0.014$,  
$\Gamma (\Xi_c^0\to \Lambda K\pi)/\Gamma (\Xi_c^0\to\Xi\pi)        = 1.07\pm 0.12 \pm 0.07$ and  
$\Gamma (\Xi_c^0\to \Lambda K^0_S)/\Gamma (\Xi_c^0\to\Xi\pi)       = 0.21\pm 0.02 \pm 0.02$
are measured for the first time.
In the decay $\Xi_c\to p K^- K^- \pi^+$, we find evidence for the 
decay $p K^- \ksbarz$  with a fractional yield of $0.51\pm 0.03\pm 0.01$.
This measurement confirms with higher precision the recent result from the CLEO collaboration~\cite{CLEO03}.

%%%%%%%%%%%%%%%%%%%%%%%%%%%%%%%%%%%%%%%%%%%%%%%%%%%%%%%%%%%%%%%%%%%%%%%%%%%%%%

\section*{Acknowledgements}

%%%%%%%%%%%%%%%%%%%%%%%%%%%%%%%%%%%%%%%%%%%%%%%%%%%%%%%%%%%%%%%%%%%%%%%%%%%%%%
We thank the KEKB group for the excellent operation of the
accelerator, the KEK Cryogenics group for the efficient
operation of the solenoid, and the KEK computer group and
the National Institute of Informatics for valuable computing
and Super-SINET network support. We acknowledge support from
the Ministry of Education, Culture, Sports, Science, and
Technology of Japan and the Japan Society for the Promotion
of Science; the Australian Research Council and the
Australian Department of Education, Science and Training;
the National Science Foundation of China under contract
No.~10175071; the Department of Science and Technology of
India; the BK21 program of the Ministry of Education of
Korea and the CHEP SRC program of the Korea Science and
Engineering Foundation; the Polish State Committee for
Scientific Research under contract No.~2P03B 01324; the
Ministry of Science and Technology of the Russian
Federation; the Ministry of Education, Science and Sport of
the Republic of Slovenia;  the Swiss National Science Foundation; the National Science Council and
the Ministry of Education of Taiwan; and the U.S.\
Department of Energy.


\begin{thebibliography}{1} 


\bibitem{PDG}   {S.~Eidelman et al., The Review of Particle Physics, 
                Phys.\ Lett.\ B 592 (2004) 1.}
\bibitem{CLEO96a} {T.~Bergfeld et al., Phys.\ Lett.\ B 365 (1996) 431.}
\bibitem{CLEO96b}{K.W.~Edwards et al., (CLEO Collaboration), 
                Phys.\ Lett.\  B 373 (1996) 261.}
\bibitem{CLEO03}{I.~Danko et al., (CLEO Collaboration), 
%                 preprint CLNS 03/1834 (2003).}
                 Phys.\ Rev.\ D 69 (2004) 052004.}
\bibitem{FOCUS01} {J.M.~Link et al., Phys.\ Lett.\ B 512 (2001) 277.}
\bibitem{FOCUS03} {J.M.~Link et al., Phys.\ Lett.\ B 571 (2003) 139.}
\bibitem{SYMQUA1} {B.~Desplanques, J.F.~Donoghue and B.R.~Holstein, 
                   Ann.\ Phys.\ (N.Y.) 124 (1980) 449.}
\bibitem{SYMQUA2} {P.~\.{Z}enczykowski,
                   Phys.\ Rev.\  D 40 (1989) 2290.}
\bibitem{ZENCZYK} {P.~\.{Z}enczykowski,
                   Phys.\ Rev.\  D 50 (1994) 402.}
\bibitem{SHARMA} {K.K.~Sharma and R.C. Verma,
                   Eur.\ Phys.\ J.\  C7 (1999) 217.}
\bibitem{KEKB}	{S.~Kurokawa and E.~Kikutani,
		Nucl.\ Instrum.\ Methods A499 (2003) 1, and other papers 
		included in this Volume.}
\bibitem{BELLE} {A.~Abashian et al., (Belle Collaboration), 
                Nucl.\ Instrum.\ Methods  A 479 (2002) 117.}
\bibitem{KALMAN} {R.E.~Kalman, Trans.\ Am.\ Soc.\ Mech.\ Eng. D. 82 (1960) 35;
		R.E.~Kalman and R.S.~Bucy., ibid.\ 83 (1961) 95.}
\bibitem{BELL}{T.~Lesiak et al., (Crystal Ball Collaboration), 
                 Z.\ Phys.\ C~55 (1992) 33.}
\bibitem{CHARM} {K.~Abe et al., (Belle Collaboration), 
                note BELLE-CONF-0335 (2003).}
\bibitem{PETERSON} {C.~Peterson et al., Phys.\ Rev.\ D 27 (1983) 105.}
\end{thebibliography}
\end{document}